\documentclass[fleqn,usenatbib]{mnras}

\usepackage{newtxtext,newtxmath}

\usepackage[T1]{fontenc}

\DeclareRobustCommand{\VAN}[3]{#2}
\let\VANthebibliography\thebibliography
\def\thebibliography{\DeclareRobustCommand{\VAN}[3]{##3}\VANthebibliography}

\usepackage{multirow}
\usepackage{multicol}
\usepackage{graphicx}	
\usepackage{amsmath}	
\usepackage[dvipsnames]{xcolor}
\usepackage{soul}

\title[Proximity of exoplanets to first-order MMRs]{Proximity of exoplanets to first-order mean-motion resonances}

\author[C. Charalambous et al.]{
C. Charalambous,\thanks{E-mail: carolina.charalambous@unamur.be}
J. Teyssandier, 
and A.-S. Libert \\
naXys, Department of Mathematics, University of Namur, Rue de Bruxelles 61, 5000 Namur, Belgium 
}

\date{Accepted XXX. Received YYY; in original form ZZZ}

\pubyear{2021}

\begin{document}
\label{firstpage}
\pagerange{\pageref{firstpage}--\pageref{lastpage}}
\maketitle

\begin{abstract}
Planetary formation theories and, more specifically, migration models predict that planets can be captured in mean-motion resonances (MMRs) during the disc phase. The distribution of period ratios between adjacent planets shows an accumulation in the vicinity of the resonance, which is not centred on the nominal resonance but instead presents an offset slightly exterior to it. Here we extend on previous works by thoroughly exploring the effect of different disc and planet parameters on the resonance offset during the disc migration phase. The dynamical study is carried out for several first-order MMRs and for both low-mass Earth-like planets undergoing type-I migration and giant planets evolving under type-II migration. We find that the offset varies with time during the migration of the two-planet system along the apsidal corotation resonance family. The departure from the nominal resonance increases for higher planetary masses and stronger eccentricity damping. In the Earth to super-Earth regime, we find offset values in agreement with the observations when using a sophisticated modelling for the planet-disc interactions, where the damping timescale depends on the eccentricity. This dependence causes a feedback which induces an increase of the resonance offsets. Regarding giant planets, the offsets of detected planet pairs are well reproduced with a classical $K$-factor prescription for the planet-disc interactions when the eccentricity damping rate remains low to moderate. In both regimes, eccentricities are in agreement with the observations too. As a result, planet-disc interactions provide a generic channel to generate the offsets found in the observations.  
\end{abstract}




\section{Introduction}
\label{sect:intro}

During the late stage of planetary formation, interactions between the planets and the gas disc in which they are embedded produce a transfer of angular momentum, changing the radial distance of the planets to the star \citep[mostly inward migration, see][]{1979ApJ...233..857G,1986ApJ...309..846L,1997Icar..126..261W}. The torque acting on the planet depends on its mass and the disc's features. Low-mass planets are affected by type-I migration, while larger planets, massive enough to open a gap in the disc, undergo type-II migration. 
More specifically, an exchange of angular momentum between the disc and the planet occurs in the vicinity of the planet's orbit, causing material to move away from the planet. If the planet is massive enough, the angular momentum exchange caused by its tidal torque can balance out the viscous flow of angular momentum, causing the creation of a gap. The planet becomes locked in its gap and follows the viscous accretion, migrating inwards. This is known as type-II migration \citep{1986ApJ...309..846L}. On the other hand, if the planet is not massive enough, viscous accretion always resupplies gas, and no gap is formed. In this regime, the small planet launches density waves in the disc at Lindblad resonances, and experiences a gravitational torque from the resulting perturbation of the disc surface density. This causes the planet to migrate, in the so-called type-I regime \citep{1997Icar..126..261W}. In most cases, the imbalance between the torque from the inner disc and from the outer disc leads to inward migration. Analytical and numerical calculations have been carried out to characterise the critical mass for which planets open a gap in the disc, indicating the transition between type-I and type-II migration. For example \citet{2003ApJ...588..494M} found that for a disc aspect ratio of 0.03, kinematic viscosity $\nu=10^{-5}$ and solar-mass stars, planets below $\sim 25 \, {\rm m}_\oplus$ are in the type-I regime, while planets larger than $\sim 100 \, {\rm m}_\oplus$ are in the type-II regime, with a transitional regime in-between.
In the following, planets with masses below $25 \, {\rm m}_\oplus$ will be denoted {\it low-mass planets}, while {\it high-mass planets} will be used for planetary masses above $100 \, {\rm m}_\oplus$ ($\sim 0.3 \, {\rm m_{Jup}}$).

The enormous variety of orbital architectures among the discovered extrasolar planets encouraged extensive study on the disc-driven planetary migration in order to understand how migration shapes the final orbits of planetary systems (\citealp[see][for detailed reviews]{2005astro.ph..7492A,2014prpl.conf..667B,2018haex.bookE.139N,2020arXiv200205756R}). During their migration, planets are expected to get trapped in MMRs. Two planets are said to be in a $(p+q)/p$ MMR, with $p,q \in \mathbb{N}$, if the mean motions $n_1$ and $n_2$ of the planets satisfy $(p+q)n_2 - p n_1\sim 0$ (i.e., commensurability of the orbital periods $P_1$ and $P_2$) and at least one of the associated resonant angles (e.g., $\phi_{1,2} = (p+1)\lambda_2 -p \lambda_1 - \varpi_{1,2}$ for first-order MMRs, as in this work) librate around a fixed value, with $\lambda_i$ the mean longitudes and $\varpi_i$ the longitudes of pericentres ($i=1,2$).

\begin{figure}
    \centering
    \includegraphics[width=\columnwidth]{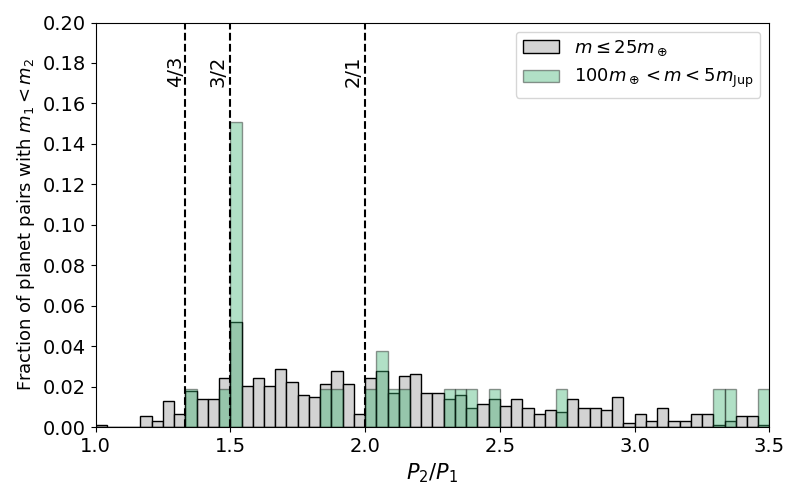}
    \caption{Normalised histogram of the orbital period ratios for pairs of planets with $m_1<m_2$. In grey, planet pairs with masses lower that $25\, {\rm m}_\oplus$, in green planet pairs with masses between $100\, {\rm m}_\oplus$ and $5 \, {\rm m_{Jup}}$. Vertical dashed lines indicate the 4/3, 3/2 and 2/1 MMRs. Most planets do not reside in (or close to) MMRs, although there is an excess of pairs with period ratio slightly larger than the nominal location of first-order MMRs (i.e., positive offset). Data from from the Extrasolar Planets Encyclopaedia (\href{http://exoplanet.eu/catalog}{http://exoplanet.eu}), as of November 2021. }
    \label{fig:histo_m}
\end{figure}

As of April 2022, almost 5000 exoplanets have been detected and confirmed. This population, especially the systems observed by the Kepler mission, shows a peak near MMRs, but not precisely at their nominal positions \citep{2014ApJ...790..146F}. The deviation of resonance can be measured with the {\it resonance offset} 
\begin{equation}
\Delta_{(p+q)/p} = P_2/P_1 - (p+q)/p.  
\end{equation}
Planets pile up at positive offsets, i.e. with orbital period ratios larger than the nominal resonant values, as shown by the period-ratio histogram of Fig.~\ref{fig:histo_m} for all the detected extrasolar planet pairs from the Extrasolar Planets Encyclopaedia \citep{2011A&A...532A..79S}, with the inner planet less massive than the outer one\footnote{This condition ensures convergent migration for planets undergoing type-I migration in the protoplanetary disc (see Section~\ref{sect:models}).} (i.e., $m_1<m_2$) and $1<P_2/P_1\le3.5$, regardless of the detection technique. The grey bins indicate pairs of planets with both masses lower than $25 \, {\rm m}_\oplus$ (low-mass planets), while the green bins indicate pairs with both masses between $100 \, {\rm m}_\oplus$ and $5 \, {\rm m_{Jup}}$ (high-mass planets). Given the greater proportion of low-mass planets, we normalised each histogram by the number of planets in each population. The distinction made here comes from the different migration regimes that the two groups of planets experience. Note also that the mass assumption $m_1<m_2$ considered here is quite representative of the observational data, since $\sim95\%$ of the detected low-mass planet pairs have the inner planet less massive than the outer one.

Fig.~\ref{fig:histo_m} shows that both the low-mass planets and the high-mass planets present an excess of positive offsets for first order MMRs compared to negative ones. More precisely, we computed the median offsets for planet pairs that lie within $\pm5\%$ of the nominal period ratios. For low-mass planets, we find $\bar{\Delta}_{2/1}=0.018$, $\bar{\Delta}_{3/2}=0.011$, and $\bar{\Delta}_{4/3}=0.009$, while $\bar{\Delta}_{2/1}=0.035$ and $\bar{\Delta}_{3/2}=0.014$ for high-mass planets (where $\bar{\Delta}$ is the median offset value of the observed exoplanet pairs whose period ratio is within 5\% of a given nominal resonance). As of April 2022, there are 98 low-mass planets within the 5\% of the nominal 2/1 MMR, 100 for the 3/2 MMR, and 30 for the 4/3 MMR, while only a few high-mass planets were detected within the 5\% of the nominal 2/1 and 3/2 MMRs. These median values suggest that high-mass planets present higher offset values than low-mass planets.  In addition, the offsets are significantly more important for the 2/1 MMR than the 3/2 MMR for both mass regimes. Let us stress that some caution has to be taken due to the uncertainties in the observations. In particular, the orbital periods of the planets detected with the radial velocity (RV) method are not precisely known, as it is the case for the transit-detected systems. The masses of the transit-detected planets for which only the radius is known are computed here following the mass-radius empirical relation of \citet{2017A&A...602A.101R}. 

Different explanations for the accumulations just outside the exact resonance were proposed. \citet{1983CeMec..30..197H,2010MNRAS.405..573P,2012ApJ...756L..11L,2013AJ....145....1B,2014A&A...570L...7D,2016AJ....152..105M,2018MNRAS.476.5032P} showed that planets can leave the commensurability following the resonance capture during the disc phase through a dissipative mechanism such as tidal interactions with the central star. However, star-planet tidal interactions are unlikely to cause divergent evolution for planets with orbital periods exceeding 10 days, as shown by \citet{2013ApJ...778....7B} who privileged divergent evolution due to wake-planet interactions during the protoplanetary disc phase. \citet{2013ApJ...770...24P} suggested that the period-ratio distribution near resonance can result from in-situ formation (\citealp[without orbital migration or dissipation, see also][]{2020MNRAS.495.4192C}). Scattering with the residual planetesimal disc after gas dispersal was also proposed to disrupt the MMR between the planets \citep{2015ApJ...803...33C}. 
\citet{2017AJ....154..236W} explored how the formation of near-resonant systems is affected by the planetary mass accretion process and the outward orbital migration of the planet pair. 

Several authors studied the possibility that the systems in the peaks just wide of the exact commensurabilities are actually resonant and formed by convergent migration. \citet{2014AJ....147...32G} showed that the direction of the shift of the peaks associated to first-order resonances to larger values than the exact resonance can be explained by the asymmetry required by convergent migration for resonance capture, while the magnitude of the offset produced by convergent migration is generally too small to match the observations of the Kepler systems.
Considering the coupling between the dissipative semi-major axis evolution and the damping of the eccentricities, they showed that for most of the Kepler systems in first-order resonance, the resonance capture could only be temporary. However, as shown by \citet{2015ApJ...810..119D}, the timescale for escape from resonance is long enough so that the intrinsic instability of the resonance cannot explain the lack of period commensurabilities within the Kepler systems. \citet{2017A&A...602A.101R} explained the departure from exact commensurability as a consequence of the smooth migration of planet pairs in a laminar flared disc. They studied the dependence of the offset values with the orbital distance of the planets and concluded that the offset distribution of the resonant configurations formed by type-I migration is consistent with the one of the Kepler planetary systems around the 2/1 and 3/2 MMRs. Moreover, they showed that during the migration the planet pair follows the family of stationary ACR solutions \citep[apsidal corotation resonance, i.e., equilibrium solutions of the averaged equations of motion, see e.g.,][]{2003ApJ...593.1124B} characterised by the simultaneous libration of both resonant angles  \citep[also denoted family of periodic orbits at zero amplitude, see][] {2006CeMDA..95..225H}, regardless of its separation from exact resonance. In other words, as a consequence of the disc eccentricity damping, the pair departs from the nominal location of the resonance, while the resonant angles librate. \citet{2021AJ....161...77W} pursue this idea by pointing out that the depletion timescale and aspect ratio of the disc also have an influence on the deviation from exact MMR. While most of the studies on the convergent migration scenario focus on first-order resonances, captures in second-order resonances have also been investigated, for instance by \citet{2015MNRAS.449.3043X} and \cite{2017MNRAS.468.3223X}.

Since the first discoveries of giant planet systems, the link between MMR capture through convergent migration and ACR families (or families of periodic orbits) was widely investigated for giant planets undergoing type-II migration \citep[e.g.,][]{2006MNRAS.365.1160B,2010CeMDA.107....3H,2014CeMDA.119..221V}, showing how the families present evolutionary tracks for planetary systems. Several studies on the MMR capture along the families were also carried out for individual detected giant planet systems \citep[e.g.,][]{2008IAUS..249..485Z, 2015A&A...573A..94T}. Recently, \citet{2020A&A...640A..55A} suggested the dynamical analysis of the families as a tool to further constrain the orbital elements of Kepler and K2 systems, while the assumption of an ACR configuration was used to constrain the fitting process of RV signals by different authors \citep[e.g.,][]{2014MNRAS.439..673B,2020AJ....160..106H}. However, to our knowledge, no study investigates in detail the offset values reached by giant planets evolving under type-II migration in the protoplanetary disc. 

In this work, we seek to analyse the offsets of resonant systems migrating in the disc and carry out a first comparison between the offsets found in the low-mass and high-mass regimes, which are characterised by type-I and type-II migration, respectively. We examine the migration of planet pairs along the ACR families and pay particular attention to how the offsets are affected by specific variations in the physical parameters of the planets (planetary mass ratio and individual masses) and in the disc parameters (eccentricity damping rate). We extend the previous results of \citet{2017A&A...602A.101R} for low-mass planets by considering more sophisticated disc model prescriptions and highlight their impact on the magnitude of the offsets. The analysis is made for initial system configurations close to the 2/1, 3/2 and 4/3 first-order MMRs, for which the inner planet is less massive than the outer one.

The paper is structured as follows. Section~\ref{sect:models} briefly describes the type-I and type-II migration models used in our $N$-body simulations. The analysis of the resonance offsets is shown in Sections~\ref{sect:simus} and \ref{sect:t-II}, for the low-mass and high-mass planets, respectively. Finally, in Section~\ref{sect:discussion}, we summarise our main results and discuss our findings in light of the observations.

\section{Migration models}
\label{sect:models}

In this work, we consider systems made of two planets on coplanar orbits and embedded in a protoplanetary disc around a central star with mass $m_\star = 1 \, {\rm m}_\odot$. We note $m_1$ and $m_2$ the inner and outer planetary mass, respectively, with $m_\star \gg m_1,m_2$. The $N$-body simulations described in this work are performed using REBOUNDx \citep{2020MNRAS.491.2885T}, where suitable prescriptions for the disc-induced migration were added. The properties of the gas disc are described in Section~\ref{sect:disc_model}. The prescriptions for the type-I migration adopted by the low-mass planets and the type-II migration for the high-mass planets are detailed in Sections~\ref{subsect:tI} and \ref{subsect:tII}, respectively. Note that we only consider here planet pairs for which both planets migrate in the same regime (i.e., either two low-mass planets or two high-mass planets).

\subsection{The gas disc} 
\label{sect:disc_model}

The initial mass of the disc is set to $5 \, {\rm m_{Jup}}$, with an inner edge at $r_{\rm in}=0.1 \, {\rm au}$ ($\simeq$~10 days for a $1 \, {\rm m}_\odot$ star), and extends up to $r_{\rm out}=50 \, {\rm au}$. We consider a disc with surface density $\Sigma$ in which planets have already formed and which loses mass over time following the prescription of \cite{2009AIPC.1158....3M}:
\begin{equation}
    \Sigma(r) = \Sigma_0 \left(\frac{r}{r_0}\right)^{-s} \exp\left(-\frac{t}{\tau_{\rm disc}}\right),    
    \label{eq:sigma}
\end{equation}
with $\tau_{\rm disc}=10^6 \, {\rm yr}$. We fix the exponent of the power-law surface density profile to $s=1$ and $\Sigma_0$ is the value at $r_0=1 \, {\rm au}$. The disc aspect ratio is given by 
\begin{equation}
    h(r) = h_0 \left(\frac{r}{r_0}\right)^f,
    \label{eq:asp-rat}
\end{equation}
where $h_0 =0.03$ and the disc flare index $f$ is set to $0.25$. Thus, the disc scale height $H(r)=h(r)r$ is not constant with radius. 

\subsection{Type-I migration regime}
\label{subsect:tI}
An analytical formula for the orbital decay timescale $\tau_a$ was initially proposed by \cite{2002ApJ...565.1257T} who developed a linear theory of the gravitational interaction between a low-mass planet (which does not open a gap in the disc) and a three-dimensional isothermal gaseous disc.
\cite{2002ApJ...565.1257T} assumed a single planet in the system, in a non-eccentric orbit and with no inclination with respect to the disc mid-plane. Their expression for the orbital migration timescale of the \textit{i}-th planet at location $r_i$ from the star is
\begin{align}
\tau_{{a_i}_{\rm I}} = \left|\frac{a_i}{\dot{a_i}}\right| = \frac{Q_a t_{{\rm w}_i}}{h(r_i)^2},
\label{eq:tau-a}
\end{align}
where the wave timescale is given by
\begin{align}
t_{{\rm w}_i} = \frac{m_\star}{m_i} \frac{m_\star}{\Sigma(r_i) r_i^2} \frac{h(r_i)^4}{\Omega_K(r_i)} ,
\label{eq:tw}
\end{align}
with $\Sigma(r_i)$ given in Eq.~\eqref{eq:sigma} and $\Omega_K(r_i) = \sqrt{G m_\star/r_i^{3}}$ the Keplerian angular velocity. The functional form for $Q_a$ given by \cite{2002ApJ...565.1257T} is $Q_a^{-1}\simeq 2.7+1.1 s$, although numerical simulations suggest different prescriptions (\citealp{2010ApJ...724..730D}). This equation shows that the migration speed is proportional to the planet-to-star mass ratio.

An updated expression was suggested by \cite{2008A&A...482..677C} who accounted for the changes in the orbital decay due to the planet's eccentricity and inclination. In the coplanar case (inclinations set to 0), their migration rate for the type-I regime reads
\begin{equation}
    \tau_{{a_i}_{\rm I}} = \frac{Q_a t_{{\rm w}_i}}{h(r_i)^2}\left[ \frac{1 + \left(\frac{e_i}{2.25 h(r_i)}\right)^{1.2} + \left(\frac{e_i}{2.84 h(r_i)}\right)^{6} }{1 - \left(\frac{e_i}{2.02 h(r_i)}\right)^4} \right].
    \label{eq:tau_a_CN}
\end{equation}
This formula shows that the migration timescale can be significantly altered when the eccentricity $e_i$ starts to exceed the disc aspect ratio $h(r_i)$.

Regarding the eccentricity damping caused by the disc, \cite{2004ApJ...602..388T} described the circularization timescale $\tau_e$ for a low-mass planet on an elliptic orbit as
\begin{equation}
\tau_{{e_i}_{\rm I}} = \left|\frac{e_i}{\dot{e_i}}\right| = \frac{t_{{\rm w}}}{0.78}.
\label{eq:tau-e}
\end{equation}
\cite{2006A&A...450..833C} adjusted this analytical formula through N-body and 2D hydrodynamic multi-planet simulations by using an ad-hoc eccentricity damping normalisation factor $Q_e$ (estimated to 0.1). Their circularization timescale for a planet in an isothermal disc is given by
\begin{equation}
\tau_{{e_i}_{\rm I}} = Q_e \frac{t_{{\rm w}_i}}{0.78}\left[1-0.14\left(\frac{e_i}{h(r_i)}\right)^2+0.06\left(\frac{e_i}{h(r_i)}\right)^3\right].
\label{eq:tau_e_CN}
\end{equation}

\begin{figure}
    \centering
    \includegraphics[width=0.95\columnwidth]{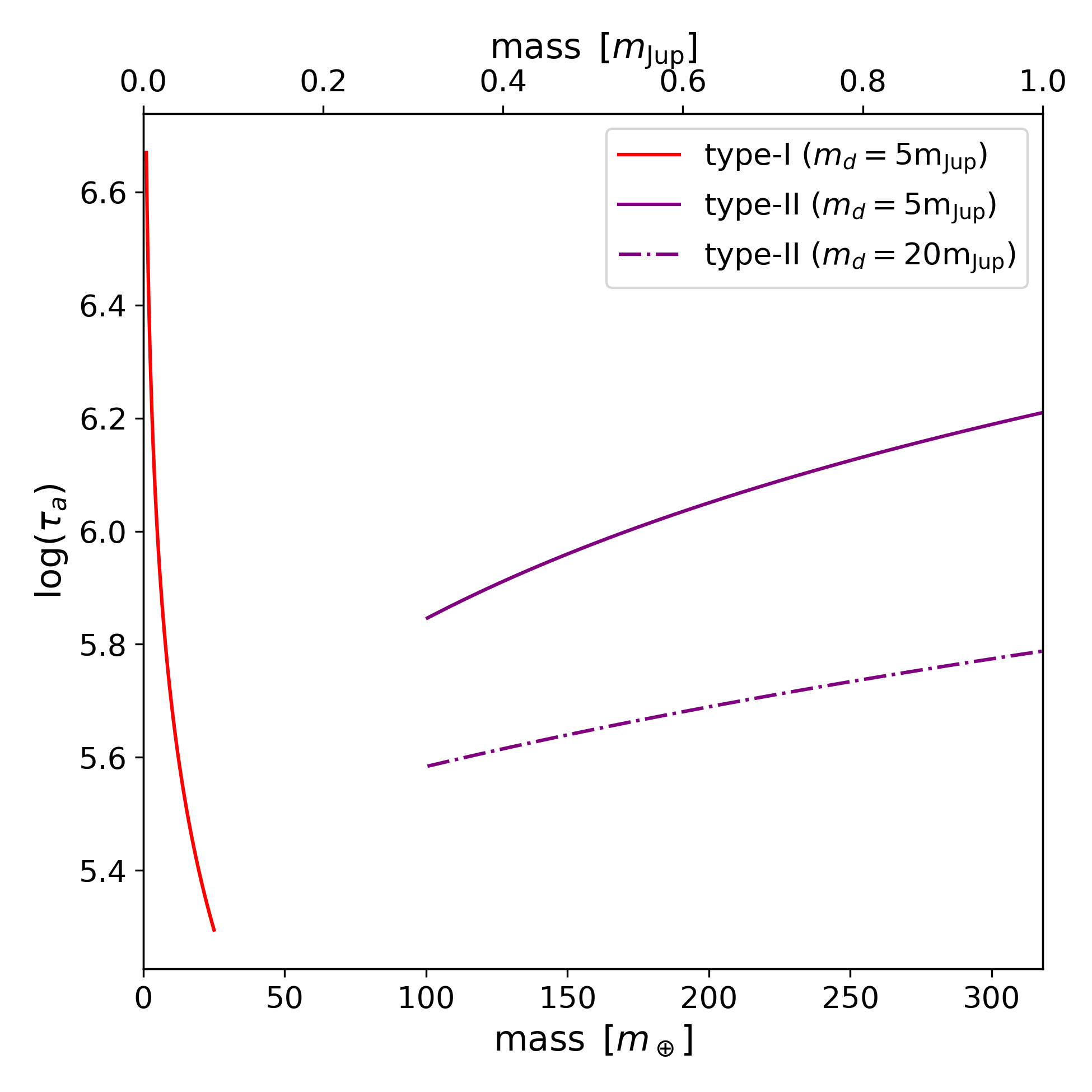}
    \caption{Migration timescale as a function of planetary mass for a planet at $a=1\, {\rm au}$ in the type-I migration regime (Eq.~\eqref{eq:tau_a_CN}, red line) and type-II migration regime (Eq.~\eqref{eq:ta_tyII}, purple lines).}
    \label{fig:mig}
\end{figure}

\begin{figure*}
    \centering
    \includegraphics[width=\textwidth]{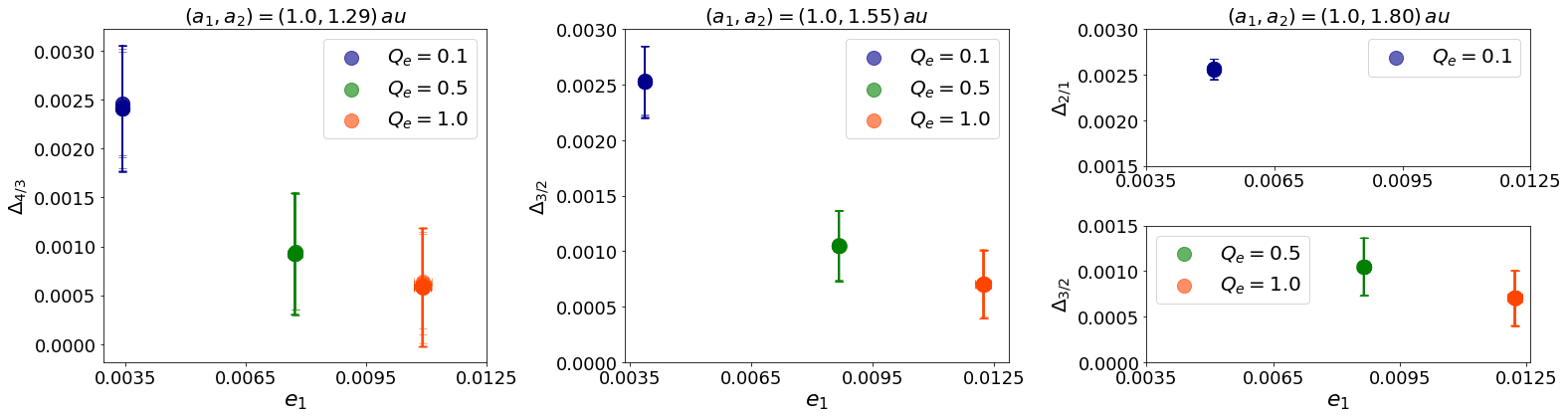}
    \caption{Resonance offsets as a function of the eccentricity of the inner planets for three different values of the $Q_e$ eccentricity damping factor of the type-I migration regime. The planetary masses are $(m_1,m_2) = (1, 3) \, {\rm m}_\oplus$. From left to right, the outer planets are initially located at $a_2=1.29, 1.55$, and $1.80 \, {\rm au}$, respectively. Circles show the median offset values and the error bars represent the variations observed during $10^5 \, {\rm yr}$ after the gas phase.}
    \label{fig:off_Qes}
\end{figure*}
\begin{figure}
    \centering
    \includegraphics[width=\columnwidth]{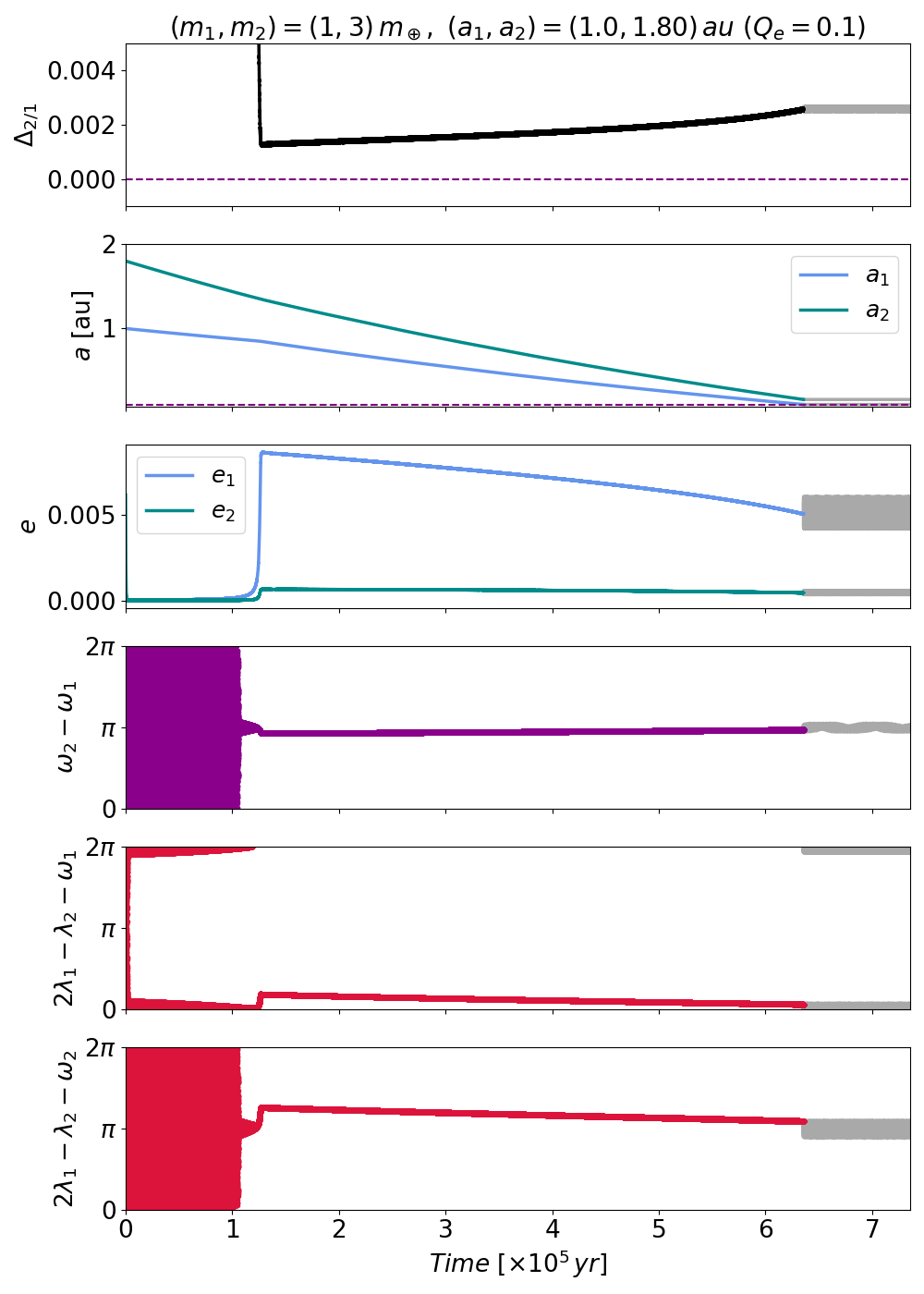}
    \caption{Time evolution of the system shown in blue in the top right panel of Fig.~\ref{fig:off_Qes} for $Q_e=0.1$. From top to bottom: the offset $\Delta_{2/1}$ (the dashed line indicates the 2/1 nominal resonance value), the semi-major axes (with a dashed line indicating the disc inner edge at $r_{\rm in}=0.1$), the eccentricities, the difference of the pericenter arguments, and the resonant angles. While the system evolves in the 2/1 MMR (anti-aligned ACRs), the offset increases with time and the eccentricities decrease simultaneously due to the damping effect of the disc. }
\label{fig:m(1,3)a(1,180)_Qe01}
\end{figure}

\subsection{Type-II migration regime}
\label{subsect:tII}
When the mass of the migrating planet grows sufficiently, it significantly alters the radial density profile, and the perturbation generated on the surrounding gas causes the formation of a gap around the planet's orbit. The surface density in the co-orbital region decreases significantly, and the drift rate is slower than in the type-I regime.

The timescale for orbital decay in type-II migration is then related to the viscous transport in the disc. Assuming a viscosity of the form $\nu=\alpha c_{\rm s}^2 \Omega_{\rm K}^{-1}$ (where $c_{\rm s}=H \Omega_K$ is the sound speed and $\alpha$ is the classical Shakura-Sunyaev viscosity parameter, which we fix at the value $\alpha = 0.001$ \citep{1973A&A....24..337S}),  then the type-II migration orbital decay is given by $a_i/\dot{a_i}=-(2/3)r_i^2/\nu$. In addition, when the mass of the planet becomes comparable to the local mass of the disc, migration slows down and eventually stops. Therefore, we follow \cite{2017A&A...598A..70S} and adopt
\begin{equation}
\tau_{{a_i}_{\rm II}} = \left|\frac{a}{\dot{a}}\right| = 
\frac{2}{3\alpha} \frac{1}{\Omega_K} \left( \frac{r_i}{H(r_i)} \right)^{2} 
{\rm Max} \left\{ 1,\frac{m_i}{(4\pi/3)\Sigma(r_i) r^2_i}\right\},
\label{eq:ta_tyII}
\end{equation}
where we consider the local surface density as the unperturbed surface density $\Sigma(r_i)$ from Eq.~\eqref{eq:sigma} \citep[see][]{2020MNRAS.492.1318S}. 
The ${\rm Max}$ function in Eq.~(\ref{eq:ta_tyII}) appears because two regimes can be distinguished within the type-II migration: the disc-dominated regime which corresponds to the classical type-II migration and occurs when the local disc mass is more massive than the planet, and the planet-dominated regime for planets more massive than the local disc mass. As previously said, in the classical type-II migration, the planet is coupled to the disc viscous evolution. Instead, when the planet is more massive than the local disc mass, the migration rate is significantly slowed down, and the gas on the outer part of the disc cannot cross the gap, causing the disc interior to the planet to viscously drain onto the star. For a disc mass of $20\, {\rm m_{Jup}}$, the disc-dominated regime takes place for planetary masses lower than $\sim 90m_\oplus$ and for a disc mass of $5\, {\rm m_{Jup}}$, it takes place for planetary masses lower than $\sim 25m_\oplus$. Hence this branch of the type-II regime is not present for the high-mass planets considered in this work ($> 100m_\oplus$). For comparison, the migration timescale is represented in Fig.~\ref{fig:mig} as a function of the planetary mass, for the type-II regime (purple curves) as well as the type-I regime (red curve).

To account for the eccentricity damping timescale in the type-II regime, we use the typical $K$-prescription \citep{2002ApJ...567..596L}, 
\begin{equation}
    \tau_{{e_i}_{\rm II}} = \frac{1}{K} \tau_{{a_i}_{\rm II}},
\label{eq:te_tyII}
\end{equation}
with $K$ a constant parameter.

\section{Offset analysis in the type-I migration regime}
\label{sect:simus}

In this section we study the disc-induced migration of two low-mass planets in the type-I regime and the resonance offset reached during the migration, for three specific first-order resonances, namely the 4/3, 3/2, and 2/1 MMRs.
\begin{figure*}
    \centering
    \includegraphics[width=\textwidth]{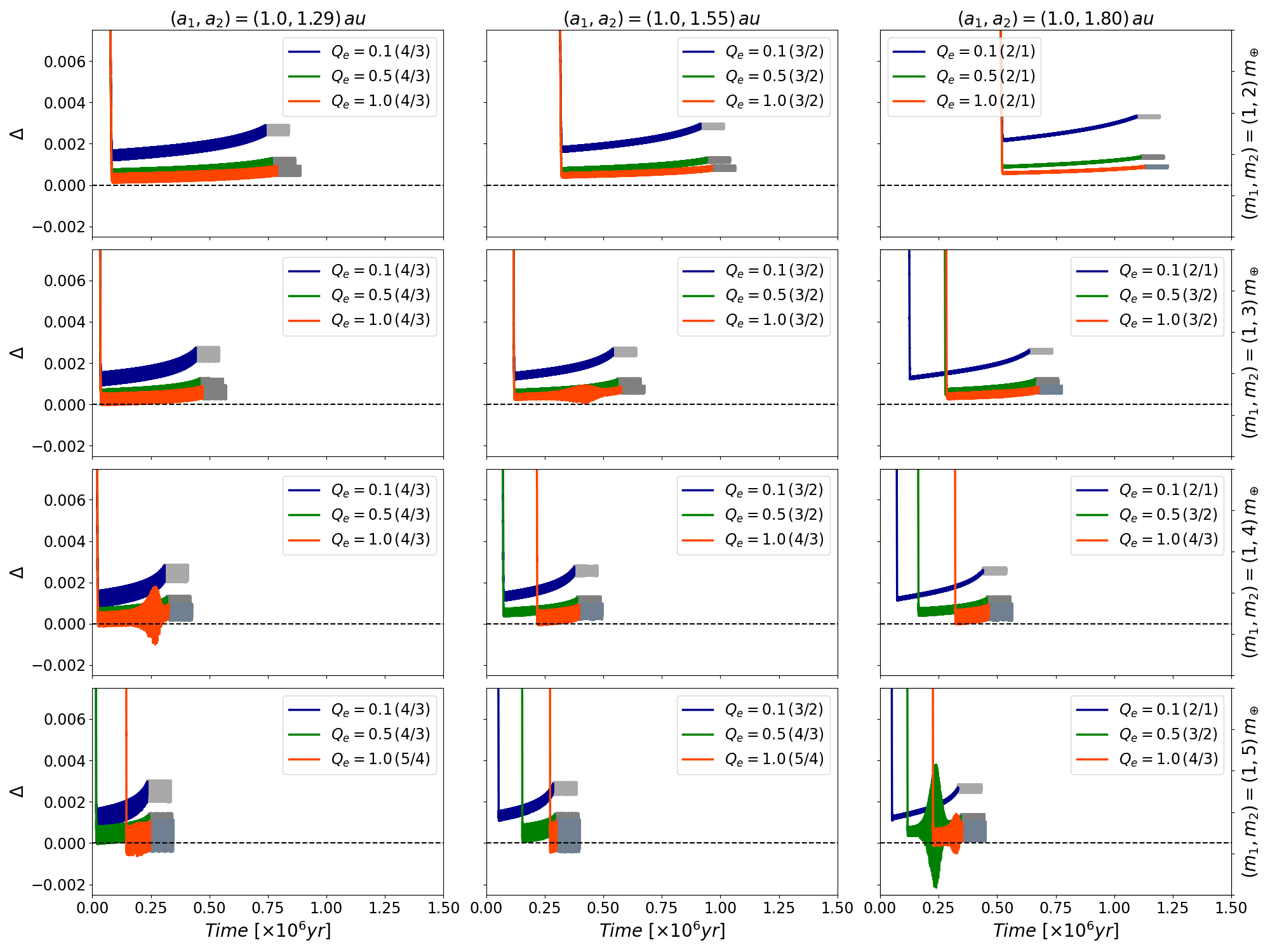}
    \caption{Resonance offsets as a function of time. Rows show decreasing mass ratio $m_1/m_2$ (see labels on the right-hand side). The different columns and colours are the same as in Fig.~\ref{fig:off_Qes}. The labels also indicate the MMR in which the systems are captured. The grey part of the plots represents the evolution $10^5$ yr after the disc dispersal. Note that for each case only one of the 10 random simulations is displayed.}
    \label{fig:Off_t}
\end{figure*}
\begin{figure}
    \centering
    \includegraphics[width=\columnwidth]{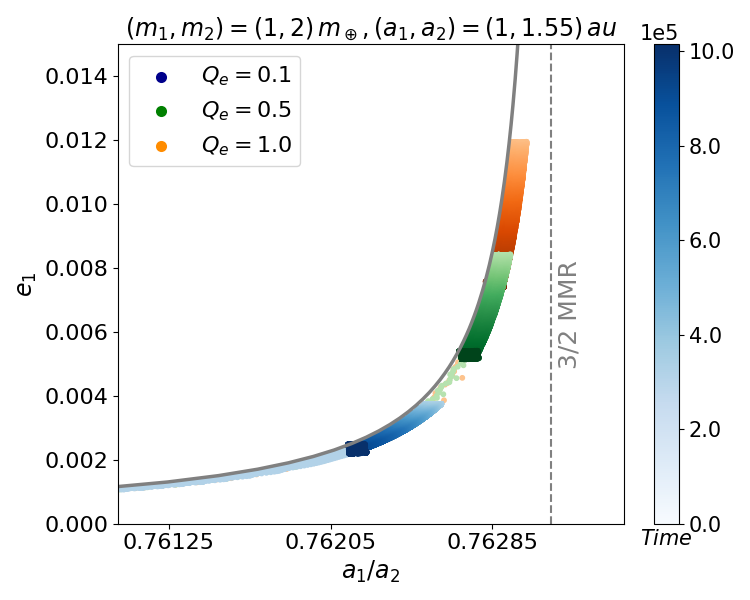}
    \caption{Time evolution of the planet pairs with $(m_1,m_2)=(1,2) {\rm m}_\oplus$ starting at $(a_1,a_2)=(1,1.55)\, {\rm au}$ in the $(a_1/a_2,e_1)$ plane, for the three values of $Q_e$. The nominal resonance is indicated in dotted line. The grey line shows the pericentric branch of the ACR family associated with the 3/2 MMR and is closely followed by the planet pairs.}
    \label{fig:ae_Qe}
\end{figure}

Systems with $m_1 \geq m_2$ do not experience convergent migration in the type-I regime (see Fig.~\ref{fig:mig}), unless the inner planet is stopped at the inner edge of the disc. In order to always guarantee convergent migration as the two planets move in the disc, we only consider systems with $m_1<m_2$ in the following. As a first step, we fix the mass of the inner planet to $1 \, {\rm m}_\oplus$ and investigate four different mass ratios by fixing the mass of the second planet to either $2,3,4$ or $5 \, {\rm m}_\oplus$. The inner planet is initially located at $a_1= 1 \,{\rm au}$ and we fix the initial position of the outer planet to $a_2= 1.29, 1.55, 1.80 \, {\rm au}$, just outside the nominal values of the considered first-order MMRs. For each of these 12 initial conditions, we select 10 random values for the other orbital parameters uniformly distributed in the intervals $e_i \in [0.001,0.01]$, and $\omega_i, M_i \in [0^\circ, 360^\circ ]$. 

The simulations for the disc-induced migration phase use Eqs.~\eqref{eq:tau_a_CN} and \eqref{eq:tau_e_CN} for the migration and circularization timescales and are stopped when $t_{\rm final}=1.5 \times 10^6$ ${\rm yr}$ or when the inner planet reaches the inner edge of the disc fixed at $r_{\rm in}=0.1\, {\rm au}$, whichever happens first. In most cases, the inner edge condition is fulfilled first. The value of $r_{\rm in}=0.1\, {\rm au}$ is in agreement with the clustering of the innermost planet of the Kepler systems around 0.1~au, as observed by \cite{2018AJ....156...24M}, which could be reminiscent of the location of the inner edge of the protoplanetary disc that acted as a planetary trap. Afterwards, we pursue the integration of the system without the disc effects for $10^5$ yr in order to study the evolution of the system after the disc dispersal. 

We study how resonance offsets are affected by the disc eccentricity damping and planetary mass ratio in Section~\ref{subsect:effect1}, as well as by the individual planetary masses in Section~\ref{subsect:effect2}. A comparison of our results with previous works using simplified prescriptions for planet-disc interactions is done in Section~\ref{subsect:tanaka}.

\subsection{Effect of the disc eccentricity damping and planetary mass ratio}
\label{subsect:effect1}

\begin{figure}
    \centering
    \includegraphics[width=\columnwidth]{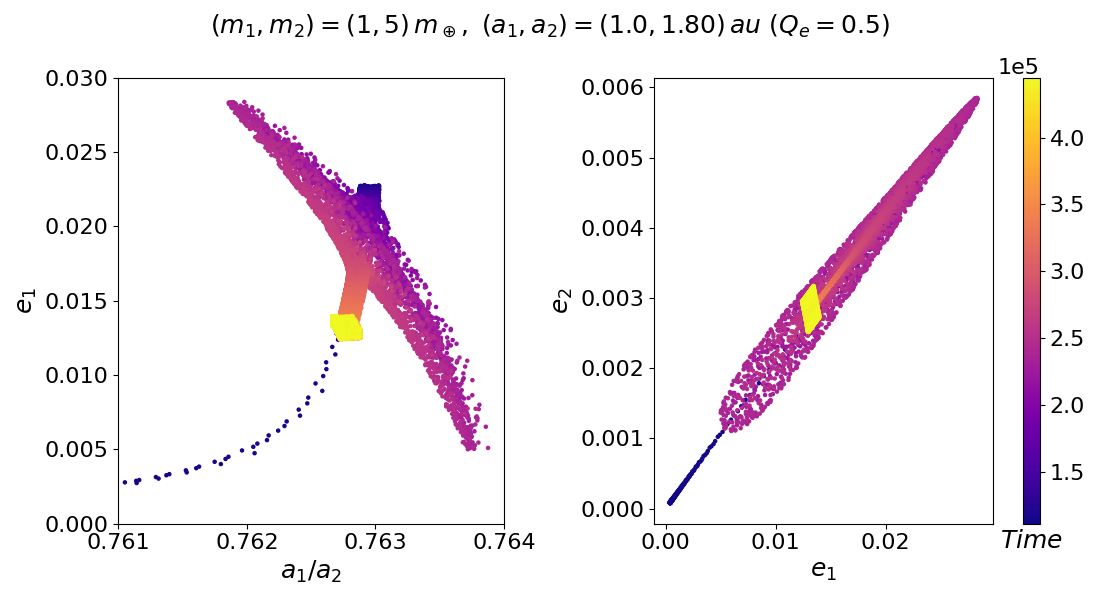}
    \includegraphics[width=\columnwidth]{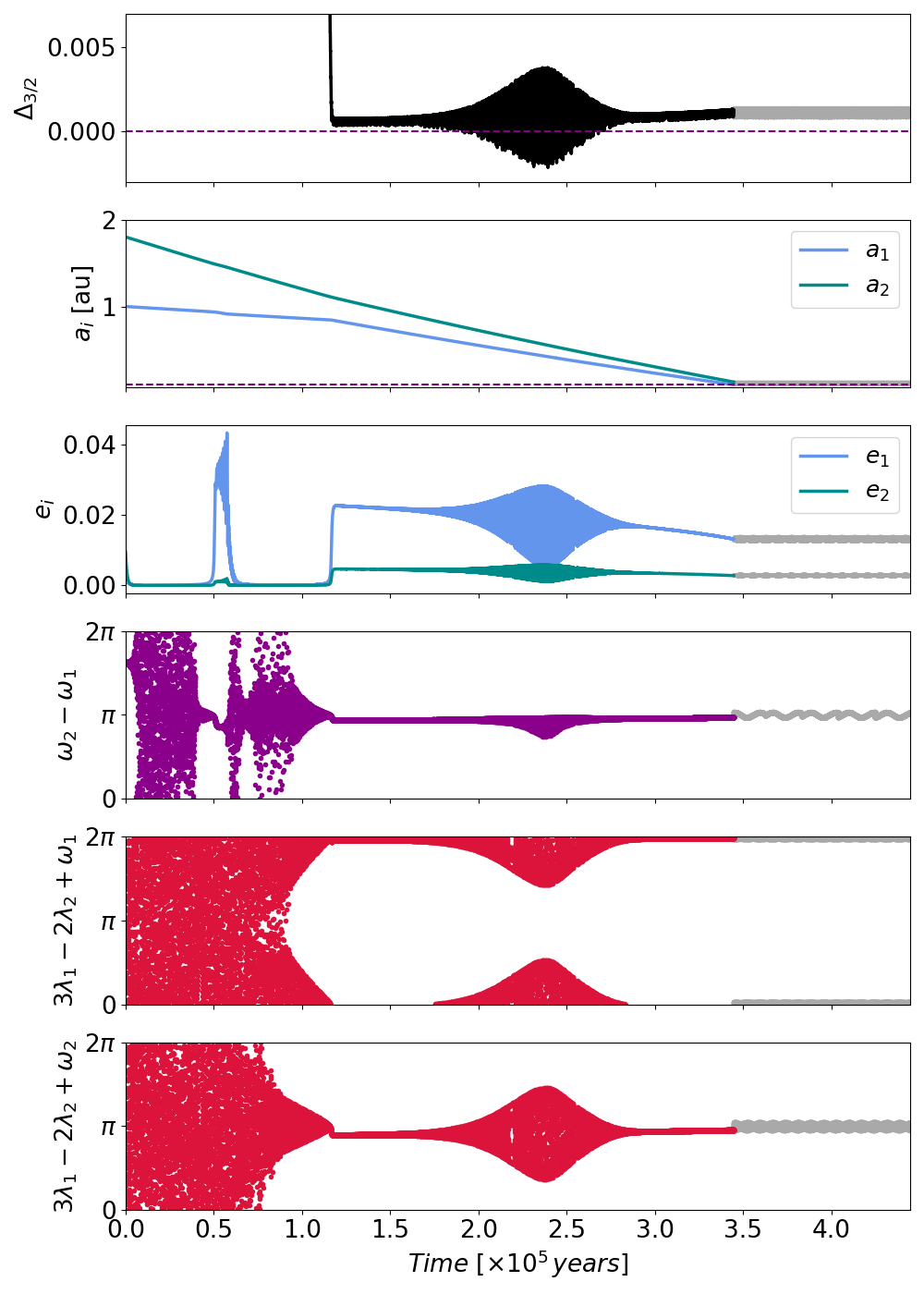}
    \caption{Top panels: Evolution of a simulation from the bottom right panel of  Fig.~\ref{fig:Off_t} with $(m_1,m_2)=(1,5) \, {\rm m}_\oplus$, starting at $(a_1,a_2)=(1,1.80)\, {\rm au}$, for $Q_e=0.5$. The colour code indicates the time evolution presented in the $(a_1/a_2,e1)$ plane in the left panel and in the $(e_1,e_2)$ plane in the right panel. The system temporarily experiences large oscillations along the ACR family associated to the 3/2 MMR. See the text for further description. Bottom six panels: Resonance offset, semi-major axes, eccentricities and resonant angles as a function of time.    }
    \label{fig:lobes}
\end{figure}

The disc parameters, which influence the migration speed and thus determine the final system architectures, still have significant uncertainties. In particular, the disc eccentricity damping is not well constrained. In the type-I migration regime this uncertainty lies in the definition of the $Q_e$ ad-hoc parameter which is fixed to 1 in the linear theory and estimated to 0.1 according to recent hydrodynamic simulations (see Eq.~\eqref{eq:tau_e_CN}). Since it was previously shown that the impact of the eccentricity damping rate is significant regarding the offset values (see Section~\ref{sect:intro}), we begin our exploration by varying the $Q_e$ parameter (i.e., the timescale it takes for a planetary orbit to be circularised by the disc) while keeping fixed all other features of the disc (as described in Section~\ref{sect:models}). Three different values will be adopted here: $Q_e=0.1, 0.5$, and $1$.

In Fig.~\ref{fig:off_Qes}, we show, for the three selected values of the eccentricity damping factor $Q_e$, the resonance offsets $\Delta$ reached by pairs of planets with $(m_1,m_2) = (1 ,3) \, {\rm m}_\oplus$ initially located just outside the different first-order MMRs, namely from left to right, the 4/3, 3/2, and 2/1 MMRs. The different $Q_e$ values will be displayed by the same colours throughout this work: $Q_e=0.1$ in blue, $Q_e=0.5$ in green, and $Q_e=1.0$ in orange. Circles show the offset values found at the end of the simulations, and the vertical error bars indicate the minimum and maximum offset values reached during the evolution of the system for the additional timescale of $10^5$ yr after the disc dispersal.

Concerning resonance capture, the systems starting just outside the 4/3 (left panel) and 3/2 (middle panel) MMRs are permanently trapped in their respective resonances for all the $Q_e$ values considered here. However, as shown in the right panel of Fig.~\ref{fig:off_Qes}, this is not the case for the systems initially close to the 2/1 MMR for which varying the $Q_e$ parameter allows the systems to get caught in different MMRs. When $Q_e=0.1$ the systems go straight to the 2/1 MMR while for $Q_e=0.5$ and $Q_e=1.0$ (i.e., for weaker eccentricity damping rates), there is a temporary capture in the 2/1 MMR but, as the migration of the planet pairs continues, the planets finally get trapped in the 3/2 MMR. As discussed previously by \citet{1993Icar..103..301B}, the resonance locking depends on the orbital elements with which the system approaches the resonance (rather than the values at the beginning of the simulation). If the disc-driven planetary migration is sufficiently slow and smooth, the orbital decay regime occurs in the adiabatic limit, the orbits remain almost circular, and it is expected that the system gets captured in the resonance, presenting low-amplitude oscillations of the resonant angles. However, even if the migration rate is sufficiently slow such that resonant capture occurs, planets can escape resonance on a timescale $\tau_e$ \citep[see eqs. 26-28 and Figure 7 from][]{2014AJ....147...32G}. This has also been observed by \citet{2021BAAA...62...41C} who showed for the TRAPPIST-1 system how varying the eccentricity damping rate of the individual planets favours captures in different MMRs.

While increasing the value of $Q_e$, or equivalently for weaker eccentricity damping rates, systems end up closer to the nominal values of the resonance. For the chosen mass ratio, the offsets vary roughly from $5 \times 10^{-4}$ for $Q_e=1$ to $2.5 \times 10^{-3}$ for $Q_e=0.1$. A similar observation was made by \citet{2017A&A...602A.101R} for the 2/1 MMR, where they used a constant orbital decay timescale and adopted a $K$-factor prescription for the eccentricity damping timescale (which is somehow inversely proportional to the $Q_e$ parameter). It is important to note that the resonance offsets $\Delta$ are always positive for the damping rates adopted here. This is due to the existence of gaps in the circular family of periodic orbits at first-order resonances which prevent the planet pair from crossing the nominal resonance in the case of adiabatic migration \citep[see for instance][for a full description of the bridges and gaps along the circular family]{2021MNRAS.tmp.1651A}. Moreover, we observe in Fig.~\ref{fig:off_Qes} that the eccentricity of the inner planet at the end of the simulation increases with $Q_e$, as expected.

In Fig.~\ref{fig:m(1,3)a(1,180)_Qe01}, we display the evolution of a typical simulation. Planetary masses are $(m_1,m_2)=(1,3) \, {\rm m}_\oplus$, the initial semi-major axes $(a_1,a_2) = (1,1.80)\, {\rm au}$, and the eccentricity damping factor $Q_e=0.1$. The top panel shows the offset $\Delta_{2/1}$ as a function of time, and the second and third plots show the time evolution of the semi-major axes and the eccentricities of both planets, respectively. During the disc-induced migration, the system is rapidly captured in the $2/1$ MMR. As both planets continue to migrate while in resonance, the resonant interactions increase the planetary eccentricities, although the disc eccentricity damping is acting on the system. We see that both resonant angles $\phi_1$ and $\phi_2$ librate, as well as their difference $\phi_2-\phi_1 = (\varpi_2-\varpi_1) = \Delta \varpi$ (see the three bottom panels), which means that the system evolves around anti-aligned ACRs (e.g., \citealp{1993CeMDA..55...25F}) with low amplitude oscillation of the angles. The system remains in this configuration until the end of the simulation.

From the top panel, we can see that the deviation from the nominal resonance (purple dotted line) is significant, even after the disc phase. After a rapid decrease, the resonance offset increases with time, while the eccentricities decrease simultaneously due to disc damping effects. In the grey part of the curve showing the evolution of the system when the disc is no longer present, the resonance offset settles down around a fixed value with (short-term) oscillations. 

Fig.~\ref{fig:Off_t} displays the time evolution of the resonance offsets for different mass ratios, for the same three values of $Q_e$ and initial locations of the outer planets as in Fig.~\ref{fig:off_Qes}. From top to bottom, the planetary masses of the outer planets are fixed to 2, 3, 4, and 5~${\rm m}_\oplus$, respectively (the mass of the inner planet is kept fixed to $1 \, {\rm m}_\oplus$ as before). For clarity, only one of the 10 random simulations is displayed for each case. Gray lines represent the evolution after the inner planet reached $r_{\rm in} = 0.1\, {\rm au}$, the moment at which the disc is removed. Let us note that it takes less time to reach $r_{\rm in}$ for systems with low mass ratios $m_1/m_2$, so they are integrated for a smaller time span. Labels for damping rates also state the MMR in which the systems are captured. We observe that the resonance capture depends on the $Q_e$ value as previously said, but also on the mass ratio between the two planets. 
\begin{figure*}
    \centering
    \includegraphics[width=\textwidth]{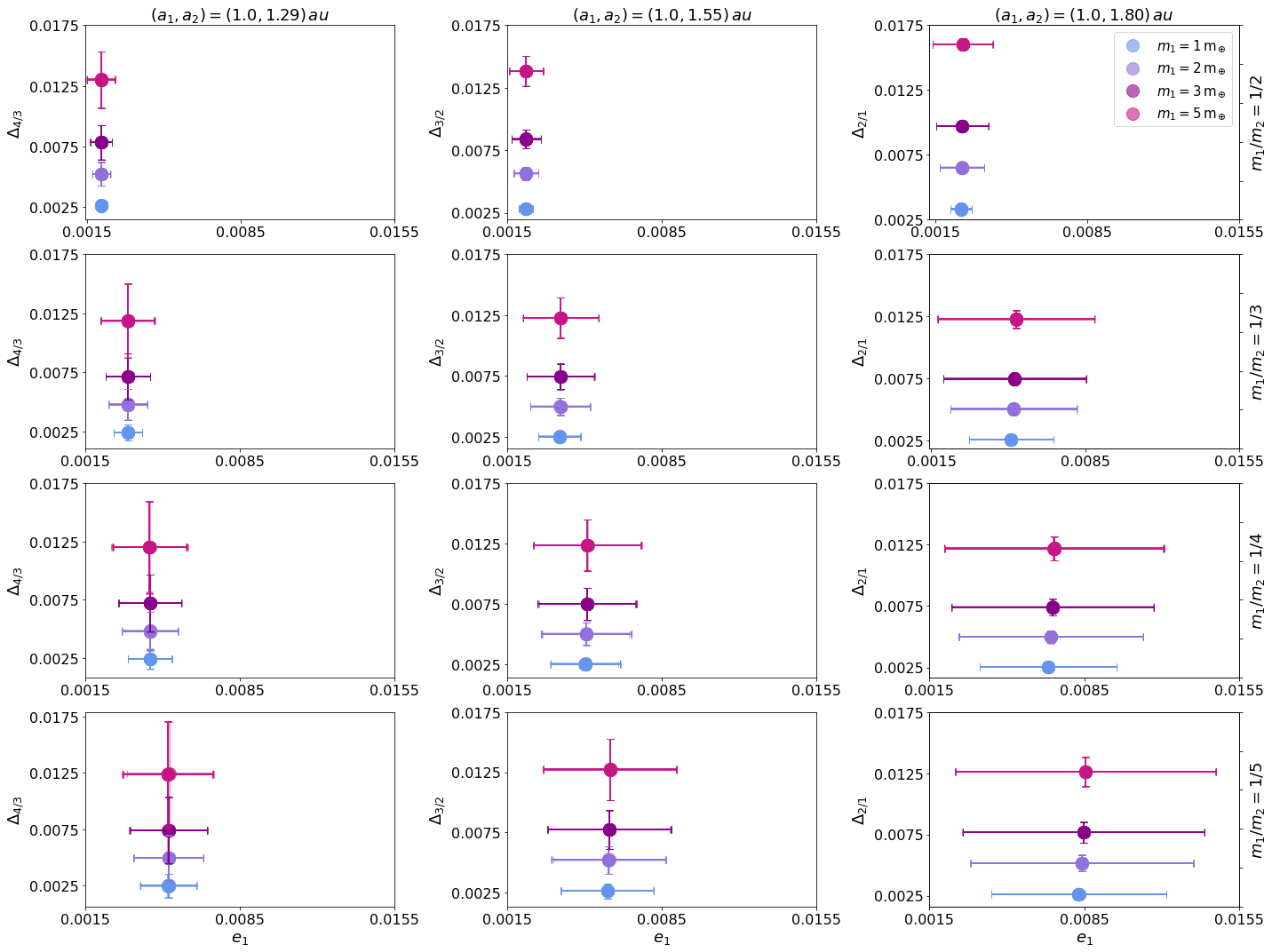}
    \caption{Resonance offsets as a function of the eccentricity of the inner planets with $Q_e=0.1$ for different individual masses (see labels in the plots and the right $y$-labels). The dots show the median values of the offsets and eccentricities for each simulation. The $y$-error bars indicate the minimum and maximum values of the offset and the $x$-error bars the ones for the inner eccentricity, during the evolution of the systems after the gas dispersal.}
    \label{fig:off_mrat}
\end{figure*}

For all the mass ratios considered, we see that the systems are locked closer to the nominal resonance value for larger $Q_e$ values. We also observe that after rapidly approaching the nominal resonance, the systems smoothly depart from it. In other words, the offset $\Delta$ starts to increase with time. Fig.~\ref{fig:Off_t} also shows that the offsets present similar values for all the mass ratios. However, the thickness of the curves indicates that the short-term variations of the offsets are more important for low mass ratios (i.e., larger outer masses) and for close locations of the planets, i.e., it varies with the strength of the resonance. The dependence of the offsets on the individual planetary masses will be investigated in Section~\ref{subsect:effect2}.

To understand better the dynamics shown in the previous figures, we report in Fig.~\ref{fig:ae_Qe} the time evolution of the planet pairs with $(m_1,m_2)=(1,2) \, {\rm m}_\oplus$ starting at $(a_1,a_2)=(1,1.55)\, {\rm au}$ (top middle panel of Fig.~\ref{fig:off_Qes}) in the $(a_1/a_2,e_1)$ plane, for the three values of $Q_e$, as well as, in grey line, the pericentric branch of the family of stationary ACR solutions associated with the 3/2 MMR which are computed from the Sessin resonant Hamiltonian (\cite{1993CeMDA..55...25F}, see also Eq.~(3) in \citealp{2017A&A...602A.101R}). The colour code refers to the time evolution, from lighter to darker colours. Although the distance to the nominal resonance (indicated in dotted line) is more significant for lower $Q_e$ and might suggest non-resonant evolutions, it is clear that all simulations closely follow the ACR family, conserving very low eccentricities, as already mentioned in \citet{2017A&A...602A.101R}. The offsets first decrease as the systems go up along the ACR family and thus approach the nominal resonance value (see also the rapid decreases in Fig.~\ref{fig:Off_t}). What is new here is that at some point, the planet pairs depart from the nominal resonance and go down along the family, due to the more sophisticated modelling for the planet-disc interactions adopted here. Indeed, this "repulsive effect" is not usually seen when using the type-I migration formula from \citet{2004ApJ...602..388T} and is a consequence of the dependence of the damping timescales on eccentricity. As the planets evolve along the ACR family, the eccentricity evolution continuously provides feedback on the damping timescales, preventing the system from reaching equilibrium. A detailed comparison when using simplified prescriptions as in the previous works will be done in Section~\ref{subsect:tanaka}. During the whole evolution, the system is locked in MMR, even for high offset values.

In some cases, we observe the appearance of `lobes' characterised by large offset variations, like in the bottom right panel of Fig.~\ref{fig:Off_t} for the systems with $(m_1,m_2)=(1,5) \, {\rm m}_\oplus$ starting at $(a_1,a_2) = (1.0, 1.80) \, {\rm au}$ for $Q_e=0.5$. This phenomenon is illustrated in Fig.~\ref{fig:lobes} with more details. The evolution of the system (from $10^5$ yr) is shown in the ($a_1/a_2,e_1$) plane in the top left panel and in the ($e_1,e_2$) plane in the top right panel. When the planet pair goes down along the ACR family, a period of large oscillations around the family affects the system during which the libration amplitude of the resonant angles increases. Then the planet pair again evolves along the family with small oscillations. For the sake of clarity, the time evolution of the system (i.e., the resonance offset, semi-major axes, and eccentricities, and the resonant angles) is also shown in the six bottom panels of Fig.~\ref{fig:lobes}. We notice that the lobes are associated with an excitation of the eccentricities, whereas the semi-major axis remain unchanged. To understand whether this effect is an artefact due to the specific prescriptions for the planet-disc interactions of \cite{2008A&A...482..677C} in the low eccentricity regime, we checked with the prescription of  \cite{2004ApJ...602..388T} (i.e., Eqs.~\eqref{eq:tau-a} and \eqref{eq:tau-e} instead of Eq.~\eqref{eq:tau_a_CN} and \eqref{eq:tau_e_CN}) and observed similar lobes. Therefore, we are confident that this phenomenon is an effect induced by the migration process. This phenomenon is possibly due to the presence of another family of periodic orbits in the proximity of the family of periodic orbits along which the system evolves. This proximity could influence the motion of the system, which then oscillates between both families and finally locks around one of them, as observed for instance in \cite{2020A&A...640A..55A}, for several Kepler and K2 systems.

\subsection{Effect of individual planetary masses}
\label{subsect:effect2}

In the previous section, we showed that the offset values are largely unaffected by the planetary mass ratio. Here we consider different individual planetary masses, keeping fixed their mass ratio. In Fig.~\ref{fig:off_mrat}, we show the same representation as in Fig.~\ref{fig:off_Qes} of the resonant offsets as a function of the eccentricity of the inner planets. The colour scale now refers to the planetary mass values. From top to bottom the rows correspond to the mass ratios $m_1/m_2 = 1/2, 1/3, 1/4$, and $1/5$. Systems with the mass of the inner planet set to $1 \, {\rm m}_\oplus$ are shown in light-blue, and the different colours are obtained by multiplying the individual masses by 2, 3, and 5. The value of the damping factor $Q_e$ is fixed at 0.1. We clearly see that the offsets are higher when the planets are more massive. In other words, the systems are found to be closer to the nominal resonance for low planetary masses. As previously discussed, the offsets vary with time. The vertical bars show, as before, the minimum and maximum offset values reached during the evolution after the dispersal of the disc. Similarly, horizontal error bars show the minimum and maximum eccentricities of the inner planets reached after the dispersal of the disc. Variations in the offset and the eccentricity increase as individual masses do. Also, the offset variations seem more important for the systems captured in the 4/3 MMR and the smallest for the systems captured in the 2/1 MMR, as previously observed in Fig.~\ref{fig:Off_t}.

The offsets that we obtained for the 4/3 and 3/2 MMRs are in the range 0.0025 to 0.013, which is similar to the observed median value for these resonances (0.009 and 0.011, respectively). Regarding the 2/1 MMR, only systems with higher individual masses approach the observed median value of 0.021. To explain this discrepancy, we first note that observations are biased towards larger planets. In order to assess this bias, we carried out additional simulations with $(m_1,m_2)=(10,20)\, {\rm m}_\oplus$ for the 2/1 MMR which show that for such super-Earth planets, the median offsets are 0.031. Therefore our ensemble of simulations can produce a large range of offsets, whose values are compatible with the observed ones. We also note that we have only explored one particular disc model and fixed the eccentricity damping factor $Q_e$ to 0.1. Other sets of parameters may lead to different offset values, but it is worthwhile to stress that the offsets found in the observations could be explained by a commonly adopted disc model approach.

Note that the range of validity of the type-I formulas for the bigger masses considered here is still debated, and the results found for them might not be as precise as in the low-mass range. In particular, the effect of radiative transfer and viscosity can prevent saturation of the corotation torque, and the negative Lindblad torques (which cause inward migration) can be cancelled by the positive contributions from the corotation region, resulting in outward migration \citep[see, e.g.,][]{2006A&A...459L..17P, 2008A&A...487L...9K}. \citet{2011A&A...536A..77B} found that this effect affects planets in the range of 20--30~${\rm m}_\oplus$.

\begin{figure*}
    \centering
    \includegraphics[width=\textwidth]{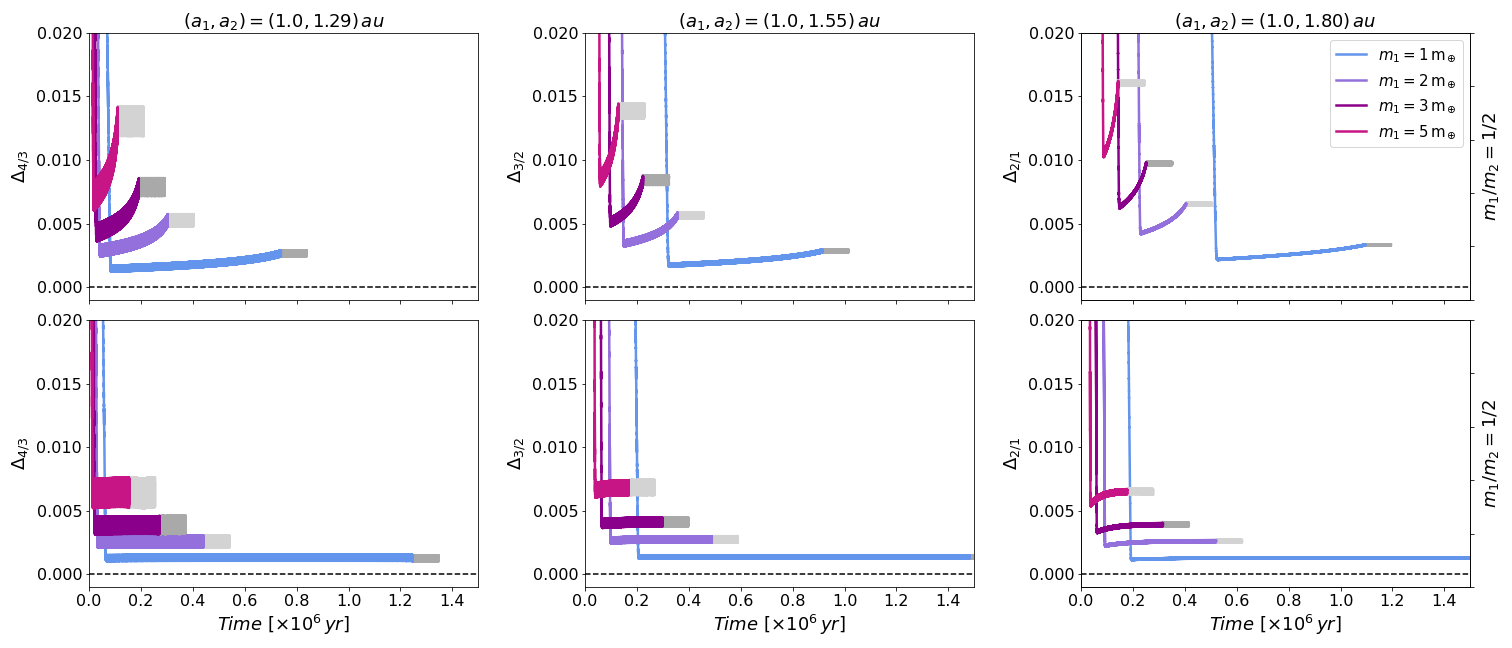} \\
    \includegraphics[width=\textwidth]{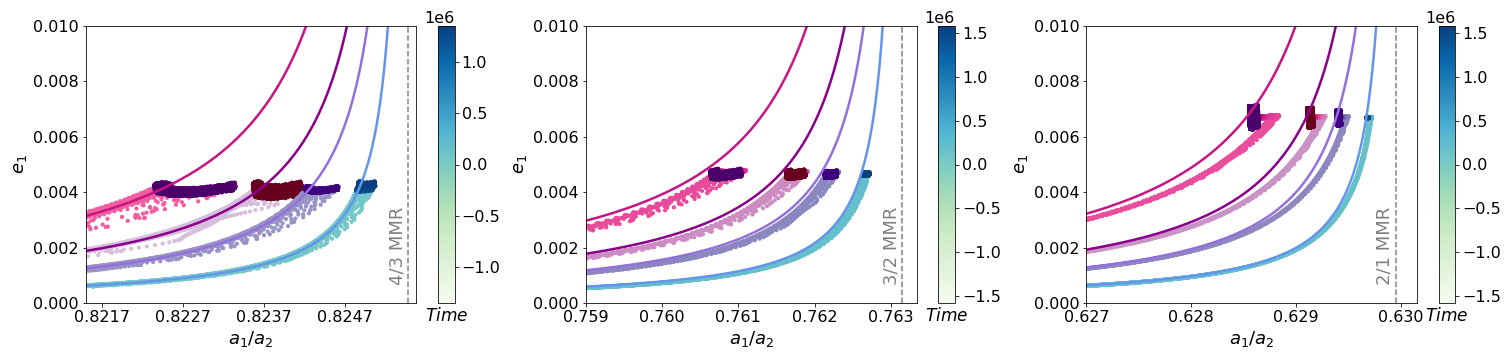}
    \caption{Top and middle panels: Resonance offsets as a function of time for the same simulations as in the top row of Fig.~\ref{fig:off_mrat} for the damping prescriptions of \citet{2008A&A...482..677C} (with $Q_e=0.1$) in the top panels and \citet{2004ApJ...602..388T} in the middle panels. Bottom panels: Time evolution of the simulations with the \citet{2004ApJ...602..388T} prescription along the pericentric branch of the ACR family corresponding to the different individual masses.}
    \label{fig:tanaka}
\end{figure*}

\subsection{Comparison with simplified damping prescription}
\label{subsect:tanaka}

In the damping prescription adopted in this work, the orbital decay and circularization timescales depend explicitly on the planetary eccentricity (see Eqs.~\eqref{eq:tau_a_CN} and \eqref{eq:tau_e_CN}). To highlight the consequences of this sophisticated prescription on the offset values, we compare our results with those obtained with the damping prescription of \citet{2004ApJ...602..388T} (Eqs.~\eqref{eq:tau-a} and \eqref{eq:tau-e}), which is independent of the planetary eccentricities. In Fig.~\ref{fig:tanaka}, the same simulations are carried out with the two different prescriptions for comparison, the \citet{2008A&A...482..677C} damping timescales in the top panels and the \citet{2004ApJ...602..388T} ones in the middle panels. The planetary mass ratio is fixed to 0.5, while the individual masses are the same as in the top row of Fig.~\ref{fig:off_mrat}, and we set $Q_e=0.1$ for the \citet{2008A&A...482..677C} prescription.

We observe that for both prescriptions, systems with more massive planets show higher offsets. However, the magnitudes of the offsets are different for the two prescriptions. The dissipative forces modelled in \citet{2004ApJ...602..388T} quickly drive the system towards an equilibrium configuration where the offset no longer varies (middle panels). Indeed, in the bottom panels showing, for each choice of masses, the time evolution of the system along the ACR family, we see that the systems go up on the ACR families, then stop, and start oscillating around the families. No subsequent increase in the offset values is observed as for the simulations with \citet{2008A&A...482..677C} prescription (top panels). With the latter prescription, as the planets evolve in the MMR along the ACR family, the eccentricity evolution continuously provides feedback on the damping times (through the terms in brackets in Eqs.~\eqref{eq:tau_a_CN} and \eqref{eq:tau_e_CN}). This feedback loop prevents the system from reaching an equilibrium and is the source of increase for the offsets with time since the planets also follow the ACR family to depart from the nominal resonance. As a result, the offsets reach higher values when considering the \citet{2008A&A...482..677C} prescription.

\section{Offset analysis in the type-II migration regime}
\label{sect:t-II}
\begin{figure*}
    \centering
    \includegraphics[width=\textwidth]{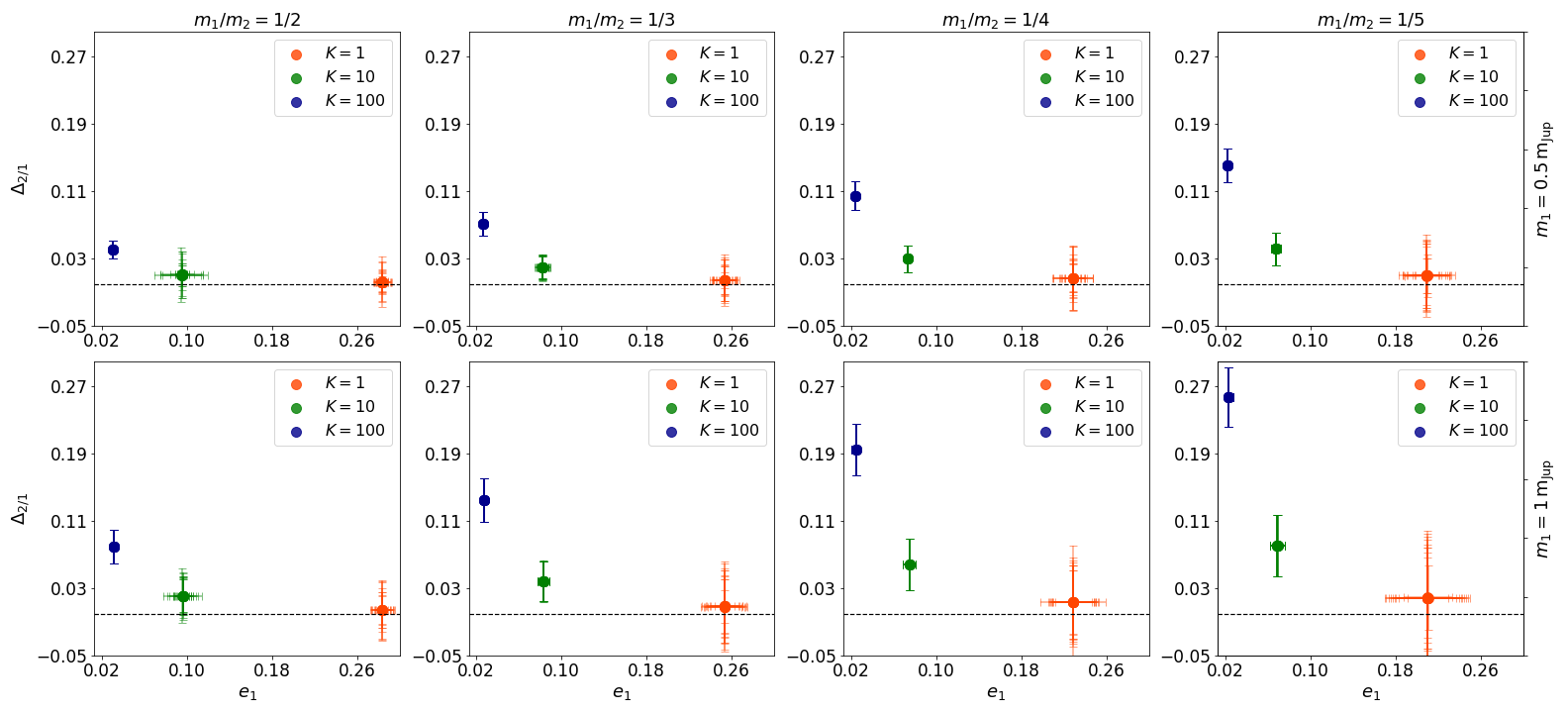}
    \caption{Resonance offsets as a function of the eccentricity of the inner planets, for three different values of the $K$-factor of the eccentricity damping prescription for the type-II migration regime. The different plots refer to different individual masses (see the labels). Eccentricity damping effects are applied both to the inner and outer planets. The circles show the median $e_1$ and median offset value for each simulation. The error bars refer to the extreme values reached during the last $10^5$ years of integration after the disc dispersal.}
    \label{fig:off_K}
\end{figure*}

\begin{figure}
    \centering
    \includegraphics[width=\columnwidth]{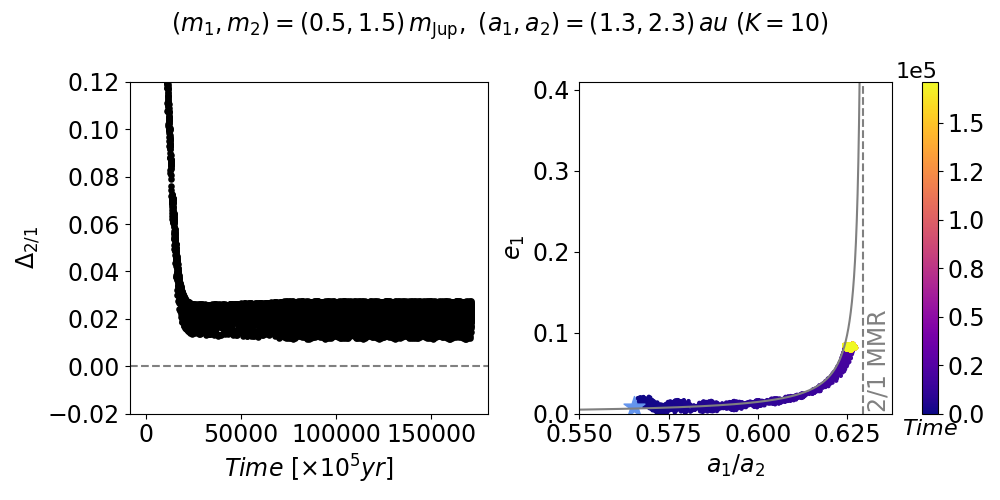}\\
    \includegraphics[width=\columnwidth]{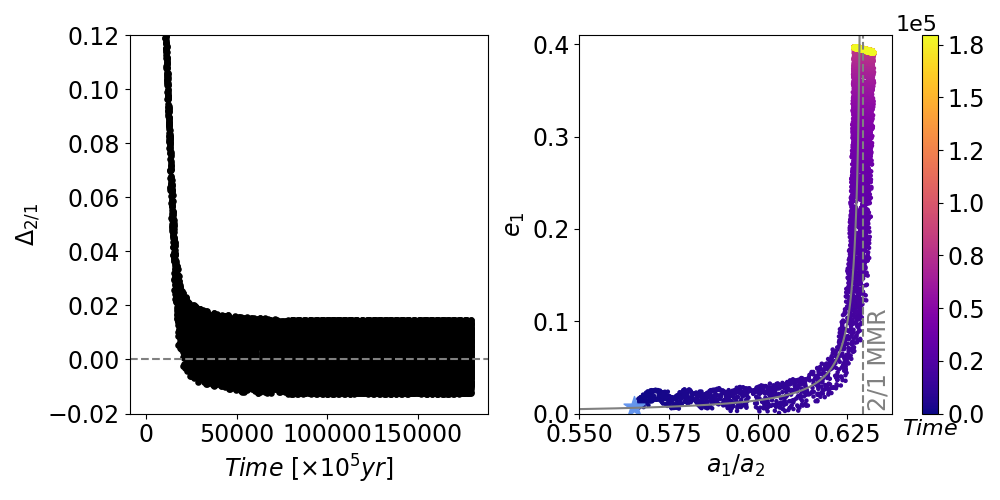}
    \caption{Time evolution of a system in the type-II migration regime with $(m_1,m_2)=(0.50,1.5) \, {\rm m_{Jup}}$ and $K=10$. In the top panels the integration includes eccentricity damping on both planets, while the bottom panels show the case where the eccentricity damping is only applied to the outer planet.}
    \label{fig:tyII}
\end{figure}

\begin{figure*}
    \centering
    \includegraphics[width=\textwidth]{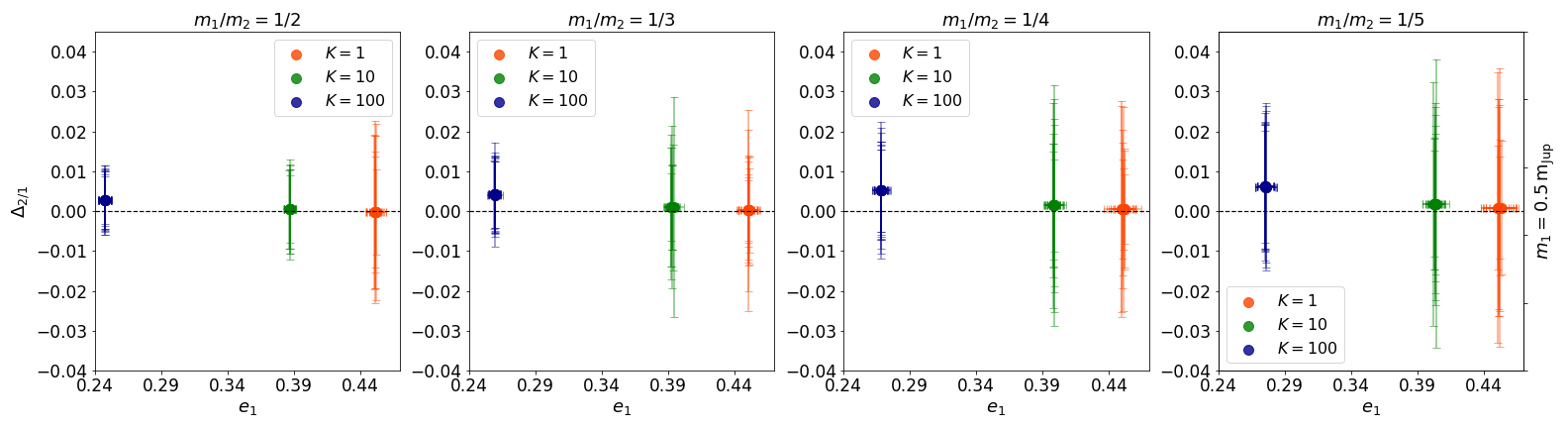}
    \caption{Same as Fig.~\ref{fig:off_K} (top panels) when considering the eccentricity damping on the outer planet only. }
    \label{fig:off_te_innerpla}
\end{figure*}

In this section we now turn to the study of the resonance offsets generated by the type-II migration regime for a pair of high-mass planets (i.e., giant-like planets with a mass higher than $100 \, {\rm m}_\oplus \sim 0.3 {\rm m_{Jup}}$). For the sake of brevity, we only focus here on the 2/1 MMR, but our results are easily transposable to other first-order MMRs. In order to produce a meaningful comparison with the previous section on the type-I regime, we need to adapt our simulations. From Fig.~\ref{fig:mig}, we see that only planet pairs with an inner more massive planet undergo convergent migration. However, to keep the same assumption about the planetary masses, namely $m_1<m_2$, we consider here, as in many studies, that the inner planet is close to the inner edge of the disc and apply an inward migration to the outer planet only (to favour convergent migration), while the eccentricity damping due to the disc is applied to both planets. We fix the inner edge of the disc at $r_{\rm in}=1$ au, since giant planets are more commonly expected to be in orbit beyond $1 \, {\rm au}$ than in close-in orbit \citep{2020MNRAS.492..377W}, and the initial positions of the planets are $(a_1,a_2)=(1.3,2.3) \, {\rm au}$.

We compute the migration of a pair of planets under the damping rates given by Eqs.~\eqref{eq:ta_tyII} and \eqref{eq:te_tyII}. Two mass values are considered for the inner planet, $m_1=0.5\, {\rm m_{Jup}}$ and $m_1=1\, {\rm m_{Jup}}$, while we adopt four different mass ratios $m_1/m_2=1/2, 1/3, 1/4$, and $1/5$. Thus, the outer planetary masses range from 1 to 5 ${\rm m_{Jup}}$, in broad agreement with the observations. Fig.~\ref{fig:off_K} shows the results of these simulations, for three different values of the $K$ damping factor in Eq.~\eqref{eq:te_tyII}: $K=1$ (orange), $K=10$ (green), and $K=100$ (blue).

As for the type-I regime, different damping timescales lead to different offset values. Higher median values of the offset are observed for higher values of the $K$-factor (i.e., for stronger damping rates), similar to what was observed for low-mass planets in the type-I migration regime. However, in the type-II regime, the ranges of oscillation for the offset during the $10^5$ yr after disc dispersal ($y$-error bars in Fig.~\ref{fig:off_K}) are broad and commonly allow the offset to take both positive and negative values, especially when $K=1$ (weak eccentricity damping). Note also that for the same $K$ value, the median offsets reached by systems with low inner planetary mass are lower than those reached when the individual masses are multiplied by two (with the same mass ratio). The same dependence on the individual masses was observed in the type-I regime. 

A typical evolution is shown in the top panels of Fig.~\ref{fig:tyII}. The parameters are fixed to $m_1=0.5 \, {\rm m_{Jup}}$, $m_2=1.5 \, {\rm m_{Jup}}$, and $K=10$. From the $(a_1/a_2,e_1)$ plane (right panel) we see that the system follows the ACR family associated to the 2/1 MMR (grey curve). Since eccentricity damping is applied on the inner planet also, the system stops his progress on the curve and presents a significant final offset close to $\Delta_{2/1}=0.02$ with limited oscillation (left panel). Let us note that unlike the simulations in the type-I regime, the value of the offset does not increase with time, departing the system from the nominal resonance. Remember that for the type-II regime, we adopt a $K$-factor prescription to model the damping on eccentricity, while the migration formulas for the type-I regime were more sophisticated (i.e., with a dependence on the planetary eccentricities too).
 
As already mentioned, in order to obtain a convergent migration we applied radial drift forces on the outer body only, while applying eccentricity damping to both planets. However, if the inner planet is inside the inner cavity, free of gas, no eccentricity damping is applied to it. In a second step, we redo the same analysis without eccentricity damping for the inner planet.  Concerning the previous system, when the eccentricity damping is applied to the outer planet only (bottom panels of Fig.~\ref{fig:tyII}), the system closely follows the family until it reaches higher eccentricity values for the inner planet ($\sim 0.4$ instead of $\sim 0.1$) and thus a closer proximity to the nominal resonance. The offset variations are large, roughly between $-0.015$ and 0.015, and they are nearly centred around the nominal resonance value.

The same simulations as in the top panels of Fig.~\ref{fig:off_K} but without eccentricity damping on the inner planet are displayed in Fig.~\ref{fig:off_te_innerpla}. As expected, the systems depart from the nominal resonance to a much lesser extent than ever before. 

For $K=1$ and 10, the period ratios show large-amplitude oscillations around the nominal resonance. These symmetric oscillations were also found in simulations by \citet{2016MNRAS.461.4406A}. Only for $K=100$ did our simulation produce noticeable positive offsets, with a median of $\sim 0.005$, still much lower than the observed median for massive planets (around 0.035). These results suggest that a model in which both planets undergo eccentricity damping produces planets with offsets closer to the observed values.

The observed distribution of period ratios for giant planets in Fig.~\ref{fig:histo_m} indicates that the massive observed exoplanet pairs have a median offset within the $5\%$ of the 2/1 MMR of $\bar{\Delta}_{2/1}=0.035$. Simulations with $K=10$ with eccentricity damping applied to both planets (green dots in the top panels of Fig.~\ref{fig:off_K}), for which the median offsets $\bar{\Delta}_{2/1}$ range from 0.01 to 0.042, are thus in good agreement with the observed population.

As an example, we consider the planetary system around GJ 876. GJ 876 b and c are two giant planets with a mass ratio of about 1/3 and a positive offset from the 2/1 MMR of about 0.037 \citep{2016MNRAS.455.2484N}. Given that the inner planet has a mass of $0.7142~m_{\rm Jup}$, this system falls between the two cases that we present on Fig.~\ref{fig:off_K}. Extrapolating from this figure, we can argue that the observed offset is compatible with disc migration in a disc with $K=10$. 

The main difference in the results observed for the low-mass and high-mass planets is due to the formulas adopted for the two migration regimes. For the type-I simulations, the prescription given by \citet{2008A&A...482..677C} is more sophisticated than the $K$-factor approach commonly used for type-II migration \citep[e.g.,][]{2002ApJ...567..596L}. In particular, the former includes a dependence on the planetary eccentricities. This dependence increases the resonant offsets as a consequence of a feedback mechanism between the planetary orbit and the disc. The commonly adopted formula for high-mass planets does not account for the interplay between the planetary eccentricity and the disc. Results for high-mass planets undergoing type-II migration then only look similar to those of the low-mass planet pairs when using a simpler prescription  \citep[such as the one of][]{2002ApJ...565.1257T}, as shown in Section \ref{subsect:tanaka}.

\section{Discussion and conclusions}
\label{sect:discussion}

In this work, we studied the dynamics of two-planet systems, for which the inner planet is less massive than the outer one, during and after the migration process, for three first-order MMRs, namely the 4/3, 3/2, and 2/1.
Migrating planets can be captured into MMR when the net effect of their migration drives them toward one another, provided the migration speed is sufficiently slow at the time they encounter the resonance. We present the results obtained when adopting migration prescriptions for the orbital decay and circularization timescales proposed by \citet{2004ApJ...602..388T} and \citet{2008A&A...482..677C} for low-mass Earth-like planets in a laminar disc and the one of \citet{2002ApJ...567..596L} for high-mass giant-like planets.

For coplanar planets in the Earth to super-Earth regime, we showed that the systems end up closer to the nominal resonance for weaker eccentricity damping  \citep[i.e., high values of the $Q_e$ parameter in][]{2008A&A...482..677C}. Besides, the departure from nominal resonance found in our simulations is always positive (due to the shape of the ACR family associated to the resonance that the system closely follows), which can explain the predominance of detected planet pairs observed just outside the nominal resonant values (see Fig.~\ref{fig:histo_m}). It is important to stress that the offset values vary with time. They decrease rapidly after the resonance capture, before a possible increase due to the disc eccentricity damping, departing the systems from the nominal resonance. This increase is observed when using sophisticated modelling for the planet-disc interactions like the one of \citet{2008A&A...482..677C} and results from the dependence of the damping timescale on eccentricity, acting as a feedback mechanism which prevents the system from reaching an equilibrium. After the protoplanetary disc phase, the offsets still exhibit significant short-term oscillations, nearly all at positive values. We stress that the median offset values found in our simulations when using the \citet{2008A&A...482..677C} prescription with $Q_e=0.1$ (i.e., strong eccentricity damping) are in agreement with the ones of the detected low-mass planets (0.009, 0.011, and 0.021 for the 4/3, 3/2, and 2/1 MMRs, respectively), unlike the offsets given by the \citet{2004ApJ...602..388T} damping formula which are too small to match the observations. Moreover, we observed a dependence of the offsets on the planetary masses (stronger deviation from nominal resonance for higher individual masses), but not on their mass ratio.

The offset analysis of the high-mass planets showed qualitatively different results from the one of the low-mass population. In particular, after MMR capture, giant planets did not exhibit an offset increase with time due to disc eccentricity damping as low-mass planets do, due to the simpler $K$-factor prescription used to model the damping effects. When the giant planets clear out a gap large enough that gas cannot flow from the outer to the inner disc, the inner disc eventually drains via accretion onto the central star, and a cavity is formed, mostly free of gas. In this situation, we can assume that the inner planet does not suffer from eccentricity damping effect and the typical result of our simulations is large and nearly symmetric oscillations around the nominal resonance value, with small median offsets, leading to possible negative values of the offsets. On the other hand, when both planets feel the effects of the disc, simulations also show large oscillations, but not centred in the nominal resonance. This second case shows good agreement with the median offset value of the giant planets (i.e., $\bar{\Delta}_{2/1} =0.035$ within the $5\%$ of the 2/1 MMR), when the eccentricity damping rate remains low to moderate ($K \sim 1$ or $10$).

The eccentricities obtained for low-mass and high-mass planets in our study are also in agreement with the observations. The mean eccentricity of the Kepler systems is around $e_{\rm mean}=0.02$ \citep{2014ApJ...787...80H,2015MNRAS.451.2589B,2015ApJ...807...44P} and thus similar to the ones reported in our simulations for low-mass planets (see Fig.~\ref{fig:off_mrat}). Also, the eccentricity values for the giant-planet population in our simulations range from 0.02 to 0.45, while the mean eccentricity for the detected giant planets is 0.29  \citep{2006ApJ...646..505B,2007ARA&A..45..397U,2017A&A...602A.107B}.

It is important to stress several limitations of our work. First, we focused here on planet pairs for which the inner planet is less massive than the outer one, in order to favour convergent migration. Second, we stopped the simulations when the innermost planet reaches the inner edge of the disc. At that point, the outer planet would normally be still surrounded by gas and would continue to migrate inward until it also reaches the cavity. However, this effect is highly related to the representation of the transition from the disc to the cavity. More sophisticated models of halting migration at the inner edge can be found, e.g., in \citet{2012ApJ...755...74K,2021A&A...648A..69A,2021A&A...656A.115H,2022A&A...658A.170T}. Although performing a detailed analysis of this final evolution is out of the scope of this work, we should mention that in the case of high-mass planets evolving in type-II migration regime, when the inner planet's eccentricity is no longer damped, we obtain offset oscillations around the nominal resonant value, albeit with a much smaller mean value compared to when both eccentricities are damped. For low-mass (terrestrial type) planets in the type-I regime, a preliminary analysis also shows a small mean value of the offset, but with unrealistically high eccentricity values. Finally, in this work, tidal interactions with the star were not included. This effect typically increases the resonant offset values, although the strength of tidal interactions is a steeply decreasing function of the distance to the star and becomes negligible for planets further than 10--20~d \citep{2012ApJ...756L..11L}. Hence, planet-disc interactions provide a more generic channel for generating offsets. Since the disc model and the considered parameters strongly affect the offset values, more detections of planet pairs in both the low-mass and high-mass regimes could help to better constrain the migration models in the future.

\section*{Acknowledgements}
We thank the anonymous reviewer for their report that helped us improve our manuscript. 
This work is supported by the Fonds de la Recherche Scientifique - FNRS under Grant No. F.4523.20 (DYNAMITE MIS-project). The work of JT is supported by a Fonds de la Recherche Scientifique – FNRS Postdoctoral Research Fellowship. Computational resources have been provided by the Consortium des Equipements de Calcul Intensif, supported by the FNRS-FRFC, the Walloon Region, and the University of Namur (Conventions No. 2.5020.11, GEQ U.G006.15, 1610468 et RW/GEQ2016).

\section*{Data Availability}
The data underlying this article will be shared on reasonable request to the corresponding author.

\bibliographystyle{mnras}
\bibliography{references} 






\bsp	
\label{lastpage}
\end{document}